\documentclass[conference]{IEEEtran}
\IEEEoverridecommandlockouts

\usepackage{caption}

\usepackage{eucal}
\usepackage{placeins}
\usepackage{makecell}
\usepackage{svg}
\usepackage{array}
\usepackage{tabularx}
\usepackage{booktabs}
\usepackage{multirow}
\usepackage{cite}
\usepackage{amsmath,amssymb,amsfonts}
\usepackage{algorithm}
\usepackage{algpseudocode}
\usepackage{graphicx}
\usepackage{textcomp}
\usepackage{xcolor}
\def\BibTeX{{\rm B\kern-.05em{\sc i\kern-.025em b}\kern-.08em
    T\kern-.1667em\lower.7ex\hbox{E}\kern-.125emX}}

\usepackage{fancyhdr}
\usepackage{hyperref}

\fancypagestyle{firstpage}{
    \fancyhf{}

    \fancyfoot[C]{%
        \footnotesize
        \href{khaled.jebari@imt-atlantique.net}{khaled.jebari@imt-atlantique.net}
        \;\textbar\;
        \href{luiz.anetneto@imt-atlantique.fr}{luiz.anetneto@imt-atlantique.fr}
        \;\textbar\;
        \href{pyndiah.ramesh@wanadoo.fr}{pyndiah.ramesh@wanadoo.fr}
        \;\textbar\;
        \href{jl.debougrenet@imt-atlantique.fr}{jl.debougrenet@imt-atlantique.fr}
    }
}

\begin{document}


\title{Quantum Block Turbo Codes}


\author{
    \IEEEauthorblockN{
        Khaled Jebari\IEEEauthorrefmark{1},
        Luiz Anet Neto\IEEEauthorrefmark{1}\textsuperscript{,}\IEEEauthorrefmark{2},
        Ramesh Pyndiah\IEEEauthorrefmark{3},
        Jean-Louis de Bougrenet de la Tocnaye\IEEEauthorrefmark{1}
    }
    \IEEEauthorblockA{
        \IEEEauthorrefmark{1}\textit{Optics Department}, \textit{IMT Atlantique}, 655 Av. du Technopôle, 29280 Plouzané, France. \\
        \IEEEauthorrefmark{2}\textit{Lab-STICC CNRS UMR 6285}, Technopôle Brest-Iroise - CS 83818, 29238 Brest Cedex 3, France. \\
        \IEEEauthorrefmark{3}\textit{Previously with the Optics Department}, \textit{IMT Atlantique}, 655 Av. du Technopôle, 29280 Plouzané, France.
    }

}

\maketitle
\thispagestyle{firstpage}

\begin{abstract}
In the early nineties, the introduction of turbo codes revolutionized classical error correction. The idea was mainly applied on two types of codes: convolutional turbo codes  and turbo product codes. The first type of codes was adapted to quantum error correction which initiated the theory of quantum serial turbo codes.
In this paper, we present a theory for quantum block turbo codes,
the quantum analog of the second type of turbo codes. We describe their iterative decoding algorithm and simulate their performances on a depolarizing channel for different constituent codes.  
\end{abstract}


\section{Introduction}
In 1993, classical error correction witnessed a historic breakthrough with the emergence of turbo codes. Berrou’s convolutional turbo codes (CTC) \cite{397441} use two 
recursive convolutional encoders separated by an interleaver, then decoded iteratively with soft information  obtained from the observation of the received sequences. This was the first demonstration of near-Shannon limit performances for classic error correction applications and opened the way for other performing decoding approaches such as Low Density Parity Check codes (LDPC) \cite{1057683} and more recently Polar codes \cite{Arikan_2009}. 
Pyndiah’s block turbo code (BTC) \cite{705396} appeared only a few years after Berrou's CTC and uses product codes \cite{1057464}, also decoded iteratively with soft decisions on the observed samples, with the interleaving structure inherently embedded in its row/column encoding and decoding procedures. 


In quantum computing, the error correction challenge is even more pronounced, with fragile quantum states highly susceptible to decoherence and manipulation errors. Consequently, quantum error correction codes (QECC) have been identified as a fundamental prerequisite for reliable quantum computation, without which errors would rapidly accumulate and render large-scale computation infeasible. 

In this context, the stabilizer formalism \cite{gottesman1997stabilizercodesquantumerror} provides a compact way to define quantum error-correcting codes using commuting Pauli operators that specify the encoded subspace and can be measured to extract error syndromes without collapsing the encoded quantum state. Calderbank-Shor-Steane (CSS) codes \cite{Calderbank_1996,PhysRevLett.77.793} are a special type of stabilizer code built from classical linear codes that separate bit-flip and phase-flip error correction. These methods make it possible to transfer a group of 
theoretical insights from classical coding theory into the quantum setting, raising the prospect of extending 
hard and soft decoding techniques to quantum information systems \cite{Babar_2015}. Furthermore, the Pauli-to-binary isomorphism paves the way for circuit-based representation of stabilizer codes, a fundamental building block 
used to build quantum encoders.

Surface codes \cite{Fowler_2012} remain today the most experimentally mature approach \cite{GoogleQuantumAI2025SurfaceCodeThreshold} for quantum error correction, offering strong performance and high error thresholds, but with an asymptotically vanishing  code rate and overhead that typically scales as $O(d^2)$ 
per logical qubit for codes with distance $d$. In contrast, quantum LDPC (qLDPC) codes \cite{Bravyi_2024} aim for constant code rate and improved scalability, with very promising theoretical results.
However, their practical implementation and decoding remain more complex, often due to the non-local qubit connectivity constraints
inherent to this approach.

Our major contribution here is the formalization of Quantum Block Turbo Codes (QBTC), the exact analog of Pyndiah's BTC. We provide an iterative maximum likelihood (ML) decoding technique for product codes using small constituent stabilizer codes to encode rows and columns of qubits. QBTCs iteratively exchange soft information between row and column decoding to estimate a correction to the qubits at the logical or physical levels. Our method could be generalized for multi dimensional product codes and holographic codes \cite{Pastawski_2015}.


We start by demonstrating decoding performances that outperform those of minimum weight (MW) decoders for single logical qubit codes. We then increase the number of logical qubits to demonstrate the turbo decoding effect. More importantly, we demonstrate how the proposed QBTC can be used as a general quantum decoding framework, allowing operation with simpler constituent codes, maintaining a relatively high code rate and a high performance due to its turbo decoding. 


This paper is organized as follows. We start with some preliminaries on circuit-based error correction representation in Section \ref{sec:prel}. We then present QBTC's encoding and decoding structures in Section \ref{sec:qbtc}. In Section \ref{sec:simu},  we show QBTC performances over a depolarizing channel. Finally, we present some perspectives in Section \ref{sec:improv} and conclude our work in Section \ref{sec:conc}.

\section{Preliminaries}\label{sec:prel}

We detail here the circuit-based representation used in this work.
Further insights on the adopted theoretical background can be found in \cite{poulin2009quantumserialturbocodes, Chandra_2019}.

\subsection{Circuit-Based Representation}

A quantum error-correcting code encoding $k$ logical qubits is defined as a $2^k$--dimensional Hilbert subspace of 
$(\mathbb{C}^2)^{\otimes n}$ and is referred to as a code of length $n$ and rate $\frac{k}{n}$. The encoding is implemented by a unitary transformation $\mathcal{V}$ acting on $n$ physical qubits:

\begin{equation}
\mathcal{C} = \left\{ \mathcal{V}\big(|\psi\rangle \otimes |0\rangle^{\otimes (n-k)}\big) \;\middle|\; |\psi\rangle \in (\mathbb{C}^2)^{\otimes k} \right\}.
\end{equation}

To obtain an efficient description, $\mathcal{V}$ is restricted to the Clifford group, i.e., the unitary transformations that preserve the $n$-qubit Pauli group $\mathbb{G}_n$ under conjugation. We denote $\mathcal{V}^\dagger$ the conjugate transpose of $\mathcal{V}$, where $\mathcal{V}^\dagger \mathcal{V} = \mathcal{V}\mathcal{V}^\dagger = I$. Such transformations admit a compact binary representation via a $2n \times 2n$ matrix over $\mathbb{F}_2=\mathbb{Z}/2\mathbb{Z}=\{\overline{0},\overline{1}\}$, called the \emph{encoding matrix} $V$ and satisfying:

\begin{equation} \label{encod matrix}
[\mathcal{V} P \mathcal{V}^\dagger] = [P] V, \quad \forall P \in \mathbb{G}_n.
\end{equation}

\noindent where $[P]$ is the binary representation of a Pauli operator $P$, with its overall phase discarded. Thus, a Clifford operator is fully specified by the matrix $V$ and a set of phases. Stabilizer codes provide an equivalent description: the code space is the joint $+1$ eigenspace of $n-k$ independent, commuting Pauli operators $\{H_i\}$. These stabilizers can be expressed as:

\begin{equation}
H_i = \mathcal{V} Z_{k+i} \mathcal{V}^\dagger,
\end{equation}

\noindent establishing the equivalence between the unitary encoding and the stabilizer formulation.

After transmission through a Pauli channel, an error $P \in \mathbb{G}_n$ affects the encoded state. Decoding via $\mathcal{V}^\dagger$ separates the error into a logical part $L \in \mathbb{G}_k$ and a syndrome part $S \in \mathbb{G}_{n-k}$ according to:


\begin{equation}
\begin{aligned}
\mathcal{V}^\dagger P |\overline{\psi}\rangle
&= \mathcal{V}^\dagger P \mathcal{V}
\big(|\psi\rangle \otimes |0\rangle^{\otimes (n-k)}\big) \\
&= (L|\psi\rangle) \otimes (S|0\rangle^{\otimes (n-k)}).
\end{aligned}
\end{equation}


Measuring the last $n-k$ qubits gives a binary \emph{syndrome} $s(P)$. Using the symplectic product $(\star)$, each syndrome bit indicates whether $P$ commutes ($0$) or anticommutes ($1$) with the corresponding stabilizer generator.

\begin{equation}
s(P)_i = [P] \star H_i.
\end{equation}

The parity-check matrix $H \in \mathbb{F}_2^{(n-k)\times 2n}$ is formed from the stabilizers and plays a role analogous to the one in classical coding theory. It encodes the stabilizer constraints and enables error-syndrome computation. Also, the syndrome depends only on the $S_x$ component:

\begin{equation}
S_{x,i} =
\begin{cases}
X & \text{if } s_i = 1, \\
I & \text{otherwise}.
\end{cases}
\end{equation}

\begin{figure}[t]
\centering
\includegraphics[width=0.4\columnwidth]{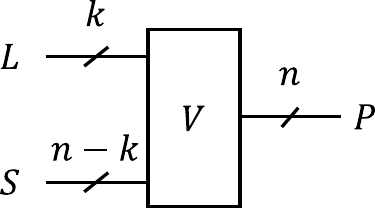}
\caption{Encoding matrix $(L:S)V=P$, defined in \cite{poulin2009quantumserialturbocodes}.}
\label{fig:encoding matrix}
\vspace{-0.6cm}
\end{figure}

Thus, two errors:

\begin{equation}
P = (L : S_x + S_z)V, \quad 
P' = (L : S_x + S'_z)V
\label{eq:degen}
\end{equation}

\noindent have the same syndrome since they differ only by $S_z$. They induce the same logical operation $L$ and can be corrected identically; such errors are \emph{degenerate}. Errors of the form:

\begin{equation}
P = (I_k : S_z)V, \quad S_z \in \{I, Z\}^{\otimes (n-k)}
\end{equation}

\noindent have zero syndrome and do not affect the logical information. A compact binary representation of the procedure is depicted in Fig.~\ref{fig:encoding matrix}.

Due to degeneracy, different physical errors may produce the same syndrome and would induce the same correcting logical operation. As a result, decoding in the quantum domain aims to identify an equivalence class (coset) of physical errors rather than an unique physical error. In that sense, maximum-likelihood decoding selects the most probable logical operator $L$ given a syndrome:

\begin{equation}
L_{\mathrm{ML}}(S_x) = \arg\max_L \ \mathbb{P}(L \mid S_x).
\end{equation}


Beyond the first key distinction from classical error correction -- namely, that the received sequence is not directly accessible -- a second fundamental difference arises in QECCs: the presence of degeneracy and the need to decode at the level of logical operators rather than individual errors.

\subsection{Quantum Encoding Matrices}

In order to construct the encoding matrix of any stabilizer code, we could start in two ways: either from the encoding circuit as described in \cite{Chandra_2019} or from the stabilizers as described in \cite{Babar_2015}. In order to simplify the representation of the encoding matrix $V$, the seed transformation $U$ is often used. Each element $U_i$ of the seed transformation:

\begin{equation}
U = \{U_1, U_2, \ldots, U_{2n}\}
\end{equation}

\noindent is defined as the decimal representation of the $i$-th row of the binary matrix $V$. 

For instance, using the first method and based on the encoding circuit of the 7--qubit Steane code presented in \cite{10049844}, we can construct the seed transformation of the $\left[\!\left[7,1,3\right]\!\right]$ code. The seed transformation of the $\left[\!\left[4,2,2\right]\!\right]$ code is given in \cite{Chandra_2019}.

Using the second method, we can construct the encoding matrix of the 5--qubit $\left[\!\left[5,1,3\right]\!\right]$ and the $\left[\!\left[8,3,3\right]\!\right]$ codes. For the 5--qubit code $\left[\!\left[5,1,3\right]\!\right]$ for instance, we have:

\vspace{-0.6cm}
\begin{equation}
V_{\left[\!\left[5,1,3\right]\!\right]} =
\begin{pmatrix}
\overline{Z}_1 \\
\overline{Z}_2 \\
\overline{Z}_3\\
\overline{Z}_4\\
\overline{Z}_5\\
\overline{X}_1\\
\overline{X}_2\\
\overline{X}_3\\
\overline{X}_4\\
\overline{X}_5
\end{pmatrix}=
\begin{pmatrix}
ZZZZZ \\
XZZXI \\
IXZZX\\
XIXZZ\\
ZXIXZ\\
XXXXX\\
IXIII\\
IIIIZ\\
IIZII\\
XIIII
\end{pmatrix}.
\end{equation}

We collected the seed transformations of these simple codes in Tab.~\ref{tab:seed codes} using the $(z \mid x)$ symplectic representation. Those will be referred to as constituent codes of our QBTC. For further details on the steps for constructing the encoding matrices, please refer to \cite{Babar_2015}, \cite{Chandra_2019} and \cite{10049844}.

In order to validate some of the constituent codes of our QBTC as well as their decoding procedure, we tested a simple maximum likelihood decoding of the $\left[\!\left[5,1,3\right]\!\right]$ and $\left[\!\left[7,1,3\right]\!\right]$ codes using degeneracy. We compared our results to the ones presented in \cite{Forlivesi_2025}, which confirmed the right choice of the encoding matrices. The performance of the 5--qubit  $\left[\!\left[5,1,3\right]\!\right]$ maximum likelihood decoder will be shown in Section \ref{sec:simu}.

\begin{table}[b]
\caption{Seed transformations of some QECCs.}
\vspace{-0.2cm}
\label{tab:seed codes}
\centering
\resizebox{\columnwidth}{!}{%
\begin{tabular}{>{\raggedright\arraybackslash}p{0.15\columnwidth}>{\raggedright\arraybackslash}p{0.75\columnwidth}}
\hline
\textbf{QECC} & \textbf{Seed transformation} \\
\hline
$\left[\!\left[5,1,3\right]\!\right]$ & \{992, 402, 201, 116, 554, 31, 8, 32, 128, 16\}$_{10}$ \\
$\left[\!\left[7,1,3\right]\!\right]$ & \{8960, 12672, 10880, 1920, 108, 90, 57, 112, 32, 16, 8, 512, 256, 128\}$_{10}$ \\
$\left[\!\left[8,3,3\right]\!\right]$ & \{21760, 13056, 3840, 255, 65280, 4005, 13226, 21910, 1283, 33797, 10274, 16384, 128, 136, 160, 192\}$_{10}$ \\
$\left[\!\left[4,2,2\right]\!\right]$ & \{144, 80, 240, 15, 10, 6, 2, 16\}$_{10}$ \\
\hline
\end{tabular}
}
\vspace{-0.5cm}
\end{table}

\section{Quantum Block Turbo codes}\label{sec:qbtc}



In this section, we present the QBTC framework that allows us to combine, in a product form, any pair of the simple codes described in the previous section. In fact, any stabilizer code whose seed transformation is known can be used as a constituent code of our QBTC. This means that QBTC gives us freedom to choose both $k$ and $n$ parameters of both codes. This section details the encoding, decoding and marginalization \& reorganization (M \& R) procedures in the QBTC.

\subsection{Encoding}
\begin{figure}[b]

\includegraphics[width=1\columnwidth]{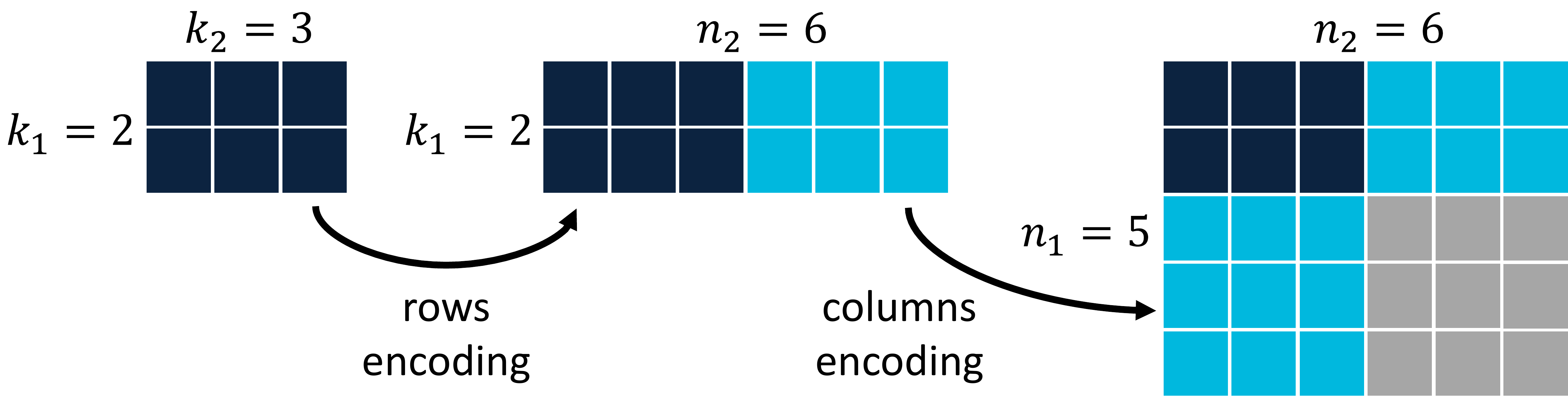}
\caption{Classical product code encoding, $(5,2) \times (6,3)$ example.}
\label{fig:product code}
\end{figure}

Throughout the remainder of this paper, we consider two codes: $C_1$ with parameters $(n_1,k_1)$ and $C_2$ with parameters $(n_2,k_2)$. 
In a classical product code $\left(n_1,k_1\right)\times\left(n_2,k_2\right)$ encoding, we would use those codes to encode $k_1\times k_2$ bits following the method described in Fig.~\ref{fig:product code}. We encode $k_1$ rows using $C_2$ then  $n_2$ columns using $C_1$. 
When systematic codes are employed, classical product codes have the property that each row and each column of the code forms a codeword of the row code and column code respectively.

This property doesn't hold, however, when using stabilizer codes because some row and column stabilizers may anticommute. For example, in the case of product code of two 5--qubit codes, the $XZZXI$ stabilizer of the first row anticommutes with the $ZXIXZ$ stabilizer of the first column. Therefore, this forces a constraint specific to this application: if we start encoding with rows then columns, we should start decoding with columns then rows and inversely.


Our code stands out for its simple
encoding scheme that uses any pair of stabilizer codes with no restrictions whatsoever regarding 
the code parameters when encoding rows then columns (or vice versa). We exploit this structure to iteratively decode while exchanging soft information between row and column decodings, as explained later in this section. Fig.~\ref{fig: encoding decoding without} shows a global view of the encoding and decoding procedures for QBTCs.


In order to encode product codes in the quantum domain, 
as shown in the upper purple box in Fig.~\ref{fig: encoding decoding without}, we start with $k_1$ rows $|\psi_1 \rangle,\dots,|\psi_{k_1} \rangle$, each with $k_2$  logical qubits. Each row, with $n_2-k_2$ ancilla qubits, gets encoded using the encoding circuit of the second constituent code. We transpose those encoded rows $|\overline{\psi_1} \rangle,\dots,|\overline{\psi_{k_1}} \rangle$ to get $n_2$ columns $|\varphi_1 \rangle,\dots,|\varphi_{n_2} \rangle$, each with $k_1$ logical qubits. Each column, with $n_1-k_1$ ancilla qubits, gets encoded using the first constituent code encoding circuit and we end up with $n_1 \times n_2$ physical qubits organized in a 2D matrix. The encoded sequence is then passed through a memoryless quantum depolarization channel. Here, we suppose a perfect measurement of the syndromes, which is a 
first step 
to pave the way for a more realistic channel model in the future. 

\subsection{Decoding} \label{sec:decoding}
\begin{figure*}[!t]
\vspace{-1cm}
\includegraphics[width=2\columnwidth]{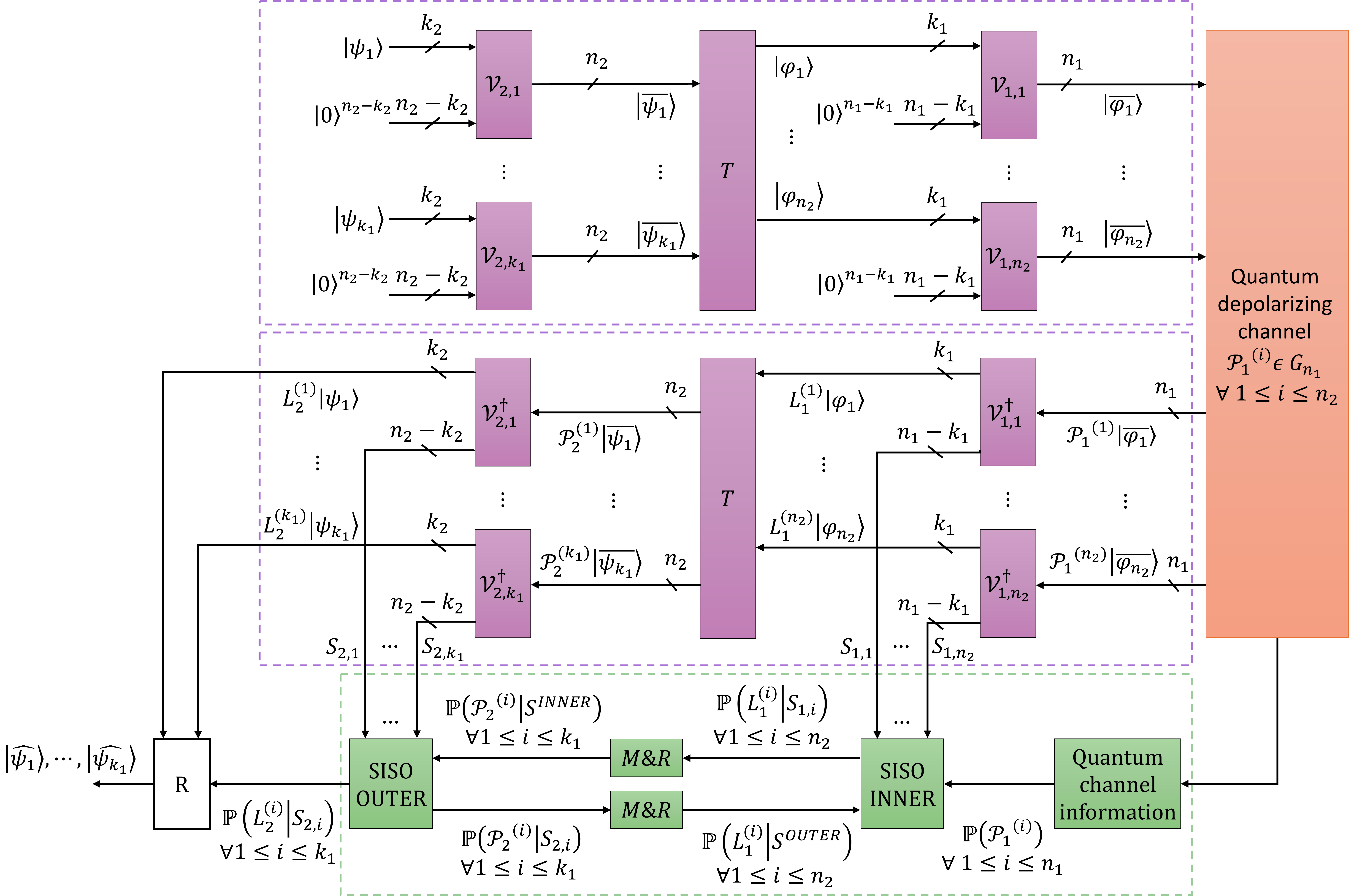}
\caption{QBTC encoding and decoding: product code $(n_1, k_1) \times (n_2, k_2)$.}
\label{fig: encoding decoding without}
\end{figure*}

Errors $\mathcal{P}^{(1)}_1,\dots,\mathcal{P}^{(n_2)}_1$ will occur in the encoded qubits $|\overline{\varphi_1} \rangle,…,|\overline{\varphi_{n_2}} \rangle$. As shown in the lower purple box in Fig.~\ref{fig: encoding decoding without}. Decoding the qubits associated to the QBTC columns will cause logical errors $L^{(1)}_1,\dots,L^{(n_2)}_1$ and ancilla errors $S^{\mathrm{INNER}}:S_{1,1},\dots,S_{1,n_2}$ on the ancilla qubits that we measure and interpret as column syndromes. Then, the logical errors on the columns are transposed to get the physical errors $\mathcal{P}^{(1)}_2,\dots,\mathcal{P}^{(k_1)}_2$ on the encoded rows, which we decode to get the logical errors $L^{(1)}_2,\dots,L^{(k_1)}_2$ on the logical qubits and ancilla errors $S^{\mathrm{OUTER}}:S_{2,1},\dots,S_{2,k_1}$ on the ancilla qubits that we measure and interpret as row syndromes. 

These syndromes along with the physical error rate of the channel are given to a decoder operating in the classic domain, which uses the exchange of soft information (soft-input, soft-output -- SISO) between the inner (SISO INNER) and outer (SISO OUTER) blocks, as shown in the green box in Fig. \ref{fig: encoding decoding without}. In order to establish an analogy with the classical block turbo codes, SISO INNER and SISO OUTER can be regarded as half-iterations whereas the M \& R procedure is equivalent to the act of transposing rows to columns and inversely after every
half-iteration. 

Each SISO block treats the syndromes of rows or columns and exchange physical and logical probabilities after transposition. We should note, however, that such transposition is not an usual matrix transposition but rather a probabilistic transposition allowing transition from row probabilities to column probabilities and back as explained later in Section \ref{sec:MR}. Based on the final logical probability after a determined number of iterations, we take a decision of an estimation of the logical errors on the logical qubits using the R block in Fig.~\ref{fig: encoding decoding without} and we apply the corresponding correction, i.e., the most likely correcting Pauli gates to the logical qubits.

\begin{figure*}[!t]
\vspace{-0.5cm}    \includegraphics[width=2\columnwidth]{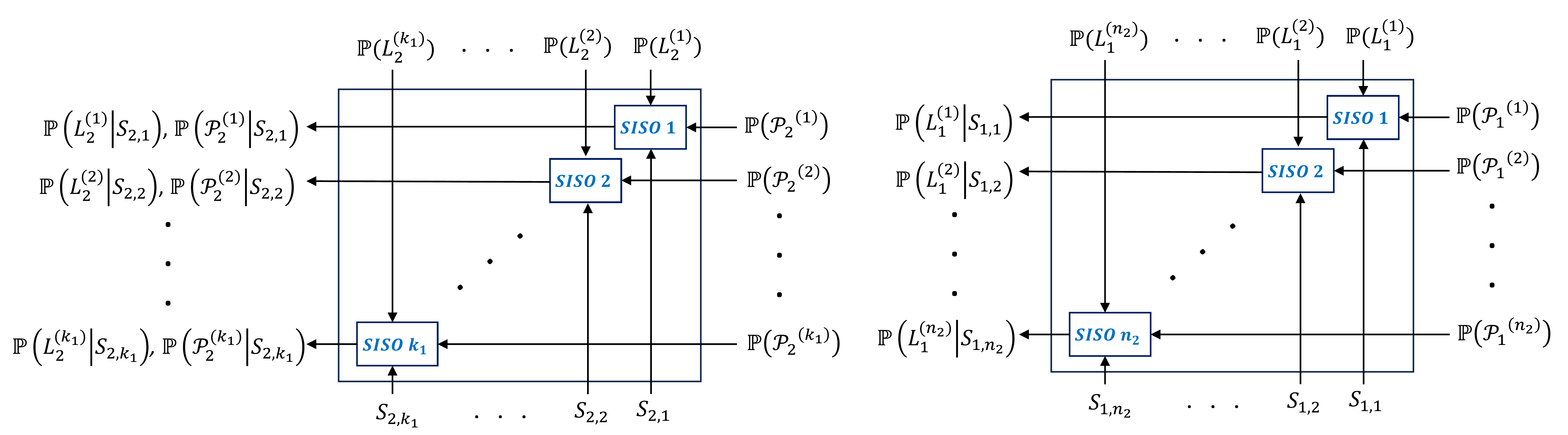}
    \caption{SISO OUTER (left) and SISO INNER (right) physical and logical probabilities.}
    \label{fig: siso without}
\end{figure*}

 The SISO INNER / SISO OUTER schemes are detailed in Fig.~\ref{fig: siso without} and will be developed in the following of this section. The SISO units inside the SISO INNER and SISO OUTER blocks obey Algorithms~\ref{alg:QBTC siso in} and~\ref{alg:QBTC siso out} respectively and can be run in parallel.

\algrenewcommand\algorithmicrequire{\textbf{Input:}}
\algrenewcommand\algorithmicensure{\textbf{Output:}}

\begin{algorithm}[t]
\caption{SISO INNER UNIT}
\label{alg:QBTC siso in}
\begin{algorithmic}

\Require $\mathbb{P}(\mathcal{P}^{(i)}_1)\  \forall 1 \leq i \leq n_2$ \quad (physical noise model)

$\mathbb{P}(L_1^{(i)})$ \quad (SISO OUTER)

$S_{1,i}$ \quad (syndrome measurement)
\Ensure 
$\mathbb{P}(L_1^{(i)} \mid S_{1,i}) , \mathbb{P}(\mathcal{P}^{(i)}_1 \mid S_{1,i})$

\medskip

\State $\mathbb{P}(L_1^{(i)} \mid S_{1,i}) \propto \mathbb{P}(L_1^{(i)})\mathbb{P}(\mathcal{P}^{(i)}_1=(L_1^{(i)}:S_{1,i})U)$

\State $\mathbb{P}(\mathcal{P}^{(i)}_1 \mid S_{1,i}) \propto \sum_{\substack{\lambda \in \mathbb{G}_{k_1} \\ \mathcal{P}^{(i)}_1=(\lambda:S_{1,i})U} }  \mathbb{P}(\mathcal{P}^{(i)}_1)\mathbb{P}(L_1^{(i)}=\lambda)$ \ 

\end{algorithmic}
\end{algorithm}

\begin{algorithm}[t]
\caption{SISO OUTER UNIT}
\label{alg:QBTC siso out}
\begin{algorithmic}

\Require $\mathbb{P}(\mathcal{P}^{(i)}_2)\ \ \forall 1 \leq i \leq k_1$ \quad (SISO INNER)

$\mathbb{P}(L_2^{(i)})$ \quad (uniform distribution)

$S_{2,i}$ \quad (syndrome measurement)
\Ensure
$\mathbb{P}(L_2^{(i)} \mid S_{2,i}), \mathbb{P}(\mathcal{P}^{(i)}_2 \mid S_{2,i})$

\medskip

\State $\mathbb{P}(L_2^{(i)} \mid S_{2,i}) \propto \mathbb{P}(L_2^{(i)})\mathbb{P}(\mathcal{P}^{(i)}_2=(L_2^{(i)}:S_{2,i})U)$
\State $\mathbb{P}(\mathcal{P}^{(i)}_2 \mid S_{2,i}) \propto \sum_{\substack{\lambda \in \mathbb{G}_{k_2} \\ \mathcal{P}^{(i)}_2=(\lambda:S_{2,i})U} }  \mathbb{P}(\mathcal{P}^{(i)}_2)\mathbb{P}(L_2^{(i)}=\lambda)$

\end{algorithmic}
\end{algorithm}

The notation used in these algorithms and the decoding in Figs.~\ref{fig: encoding decoding without} and ~\ref{fig: siso without} is denoted  as follows: $\mathbb{G}_{k}$ represents the $k$ qubits Pauli group, $\mathbb{P}(\mathcal{P}_1)$  the probability of the physical input of the inner SISO, $S_1$ the columns syndromes, $\mathbb{P}(L_1|S^{\mathrm{INNER}})$ (intermediate logical columns) the probability of the logical output of the SISO INNER. 

After transposition, we get $\mathbb{P}(\mathcal{P}_2|S^{\mathrm{INNER}})$ (intermediate physical rows) the probability of the physical input of the outer SISO. We denote $S_2$ the syndromes of rows, $\mathbb{P}(L_2|S^{\mathrm{OUTER}})$ the probability of the logical output of SISO OUTER, $\mathbb{P}(\mathcal{P}_2|S^{\mathrm{OUTER}})$ the probability of the physical output of SISO OUTER. 

When transposed again, we get $\mathbb{P}(L_1|S^{\mathrm{OUTER}})$ the probability of the logical input of SISO INNER. We denote $(L:S)$ the concatenation of the Pauli gates that affected the logical and the ancilla qubits. We represent 
the encoding by multiplying by the seed transformation $U$.

We would like to emphasize that the degeneracy described in Eq.~\ref{eq:degen} must be taken into account when calculating probabilities as follows:


\begin{equation}
\mathbb{P}(\mathcal{P}=(L:S)U)=\sum_{S_z}\mathbb{P}(\mathcal{P}=(L:S+S_z)U) 
\end{equation}

Note that if we add entanglement assistance \cite{Wilde_2014} on the last $m$ ancilla qubits $(m \leq n-k)$ we would sum over $S_{z,m}=\{I,Z\}^{\otimes (n-k-m)} \otimes \{I\}^{\otimes m}$ instead.


The initialization of the classical decoding part  is done by setting the physical probabilities to the channel's and the logical error probabilities to a uniform distribution. Looking inside an iteration and following the scheme presented in the green box in Fig. \ref{fig: encoding decoding without}, the inputs of SISO INNER are the syndromes of the columns, the physical probabilities coming from the noisy channel and the logical probabilities coming from SISO OUTER, which we assume uniform in the first iteration. Following that, SISO INNER computes and returns the updated logical probabilities which serve, after marginalization and re-organization (M \& R), as the input physical probabilities to SISO OUTER along with row syndromes and logical uniform probabilities. Subsequently, SISO OUTER computes and returns the updated logical error probabilities used to make the correction decision and the updated physical probabilities which serve, after M \& R, as the input logical probabilities of SISO INNER. We repeat this process for a fixed maximum number of iterations and apply the correction corresponding to the most probable logical error.


As shown in Algorithms~\ref{alg:QBTC siso in} and~\ref{alg:QBTC siso out}, in order to update the logical probabilities, we assume a logical Pauli sequence and we take into account all the Pauli gates that could have affected the ancillas to generate the syndrome. This procedure can be seen as some sort of quantum domain  equivalent of Chase's algorithm \cite{1054746} used in classical BTC. Next, we encode using the encoding matrix to gain access to the physical Pauli gates to which we associate its physical probability. Lastly, we sum over all the possibilities and we multiply by the probability of logical error sequence from the input. In order to update the physical probabilities, we take a physical Pauli sequence and we consider all the logical and ancilla Pauli gates combinations that could have given this specific physical sequence after encoding. Following that, we sum the logical error probabilities that verify this condition. Finally, we mutliply by the probability of the physical sequence from the input. These procedures are represented in the pseudo code in Appendix~\ref{halfiter}. 



\subsection{Marginalization \& Reorganization} \label{sec:MR}


The marginalization and reorganization procedure  consists of passing soft information between the inner and outer decoders, changing column probabilities to row probabilities and vice versa. In this section, we detail this procedure from the inner to the outer block then from the outer to the inner block. An example and pseudocode are given in Appendix \ref{Trans}.

\subsubsection{M \& R $\mid$ INNER $\rightarrow$ OUTER}

Going from the inner block to the outer block, we transform the output inner logical column error probabilities into input outer physical row probabilities. 

We start with a column represented with a sequence of $k_1$ Pauli operators:

\begin{equation}
L_1^{(i)}=P^1_i \otimes \dots \otimes P^{k_1}_i , 1\leq i \leq n_2
\end{equation}

We then marginalize the $n_2$ columns over the $k_1$ positions. In other words, $\forall \ 1 \leq j \leq k_1$:

\begin{equation}
 \mathbb{P}(P^j_i \mid S^{\mathrm{INNER}})=\sum_{P^1_i,...,P^{j-1}_i,P^{j+1}_i,...,P^{k_1}_i}\mathbb{P}(L_1^{(i)} \mid S_{1,i})
\end{equation}

A row can be seen as a sequence of $n_2$ Pauli operators:

\begin{equation}
\mathcal{P}^{(k)}_2=P^k_1 \otimes \dots \otimes P^k_{n_2} , 1\leq k \leq k_1
\end{equation}

We then reorganize these positions to have the Pauli sequences representing the $k_1$ rows:

\begin{equation}
\mathbb{P}(\mathcal{P}^{(k)}_2 \mid S^{\mathrm{INNER}})= \prod_{i=1}^{n_2} \mathbb{P}(P^k_i \mid S^{\mathrm{INNER}})
\end{equation}

In this procedure, we exchange a posteriori information. To transmit only the extrinsic information, we need to remove a priori information \cite{Wilde_2014}. We divide in the marginalization step by the marginalization of the input probabilities as follows:

\begin{equation}
 \mathbb{P}^e(P^j_i) \propto \frac{\mathbb{P}(P^j_i \mid S^{\mathrm{INNER}})}{\mathbb{P}^a(P^j_i)}
\end{equation}

\noindent with:

\begin{equation}
 \mathbb{P}^a(P^j_i)=\sum_{P^1_i,...,P^{j-1}_i,P^{j+1}_i,...,P^{k_1}_i}\mathbb{P}(L_1^{(i)})
\end{equation}

Thus:

\begin{equation}
\mathbb{P}^e(\mathcal{P}^{(k)}_2)= \prod_{i=1}^{n_2} \mathbb{P}^e(P^k_i)
\end{equation}

\subsubsection{M \& R $\mid$ OUTER $\rightarrow$ INNER}

Going from the outer block to the inner block, we transform the output outer physical row probabilities into input inner logical column error probabilities. 

We start with a row represented with a sequence of $n_2$ Pauli operators:

\begin{equation}
\mathcal{P}^{(k)}_2=P^k_1 \otimes \dots \otimes P^k_{n_2},1 \leq k \leq k_1
\end{equation}

We then marginalize the $k_1$ rows over the $n_2$ positions. Thus, $\forall \ 1 \leq i \leq n_2$:

\begin{equation}
\mathbb{P}(P^k_i \mid S^{\mathrm{OUTER}})= \sum_{P^k_1,...,P^k_{i-1},P^k_{i+1},...,P^k_{n_2}} \mathbb{P}(\mathcal{P}^{(k)}_2 \mid S_{2,k})
\end{equation}

A column can be seen as a sequence of $k_1$ Pauli operators:

\begin{equation}
L_1^{(j)}=P^1_j \otimes \dots \otimes P^{k_1}_j,1 \leq j \leq n_2
\end{equation}

We then reorganize these positions to have the Pauli sequences representing the $n_2$ columns. Therefore:

\begin{equation}
\mathbb{P}(L_1^{(j)} \mid S^{\mathrm{OUTER}})=\prod_{k=1}^{k_1} \mathbb{P}(P^k_j \mid S^{\mathrm{OUTER}})
\end{equation}

We adopt the same procedure as before to work with the extrinsic information:

\begin{equation}
\mathbb{P}^e(P^k_i) \propto \frac{\mathbb{P}(P^k_i \mid S^{\mathrm{OUTER}})}{\mathbb{P}^a(P^k_i)}
\end{equation}

\noindent with: 

\begin{equation}
\mathbb{P}^a(P^k_i)= \sum_{P^k_1,...,P^k_{i-1},P^k_{i+1},...,P^k_{n_2}} \mathbb{P}(\mathcal{P}^{(k)}_2)
\end{equation}

Thus:

\begin{equation}
\mathbb{P}^e(L_1^{(j)})=\prod_{k=1}^{k_1} \mathbb{P}^e(P^k_j)
\end{equation}

We present in Appendix~\ref{Trans}
an example of these transposition steps. We start by marginalizing each position in each column. For example, to get the probability of having the identity on the first position in the first column, we sum all the probabilities of all the Pauli sequences that obey this condition. Then, to get the rows probabilities, we multiply the probabilities of each Pauli on each position in the chosen sequence. We finally extract the extrinsic information by removing the a priori information after marginalization. 


\section{Simulations and Results} \label{sec:simu}

\begin{figure}[t]
\vspace{-0.6cm}
\includegraphics[width=\linewidth]{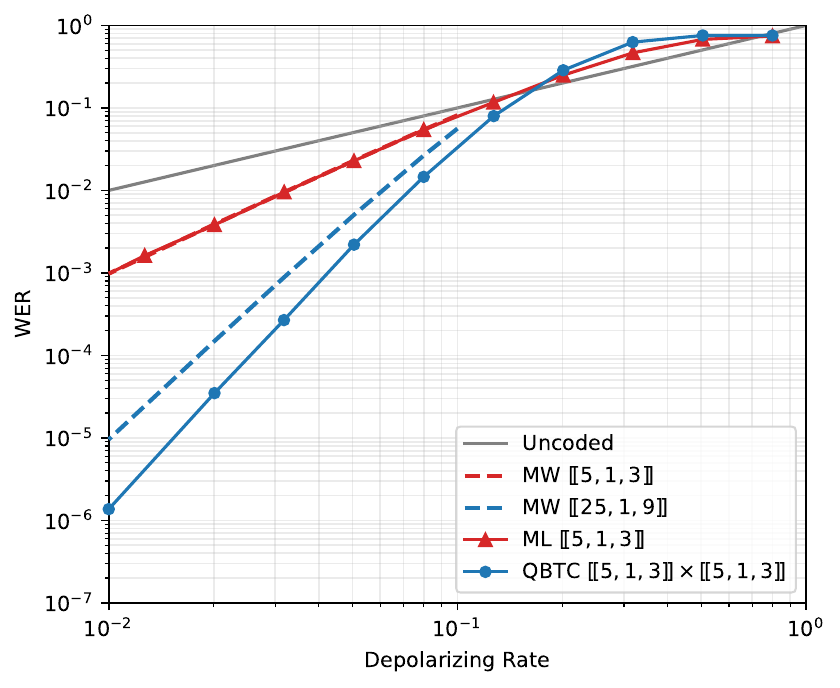}

\caption{Comparison performances of $\left[\!\left[5,1,3\right]\!\right]$ maximum likelihood (ML) decoding, $ \left[\!\left[25,1,9\right]\!\right]$ minimum weight (MW) decoding and QBTC $\left[\!\left[5,1,3\right]\!\right] \times \left[\!\left[5,1,3\right]\!\right]$.}
\vspace{-0.5cm}
\label{fig: qbtc courbe}
\end{figure}

We consider a code-capacity noise model in which each data qubit is independently subjected to a depolarizing channel with error probability $p$. Syndrome measurements are assumed perfect, i.e., the decoder receives the syndromes without measurement noise. 
Our simulations are implemented in C$++$. All Monte Carlo simulation results were obtained for at least $500$ errors were observed, unless otherwise stated.



We start by evaluating a QBTC $\left[\!\left[5,1,3\right]\!\right] \times \left[\!\left[5,1,3\right]\!\right]$, with word error rate (WER) performances depicted in Fig.~\ref{fig: qbtc courbe}. For depolarizing rate of 
\(10^{-2}\), the simulations were limited to $34$ observed errors. Those are complemented by theoretical approximations
of the word error rate, where the worst-case scenario was considered: no entanglement assistance and with a focus on an upper bound of the logical error. We also show the performances of the MW and ML decodings of the constituent $\left[\!\left[5,1,3\right]\!\right]$ code. As expected for a perfect code, we can see that the performances of the ML $\left[\!\left[5,1,3\right]\!\right]$ decoder match those of the MW approach. Those results are also in good agreement with  \cite{Forlivesi_2025}. When comparing now our QBTC $\left[\!\left[5,1,3\right]\!\right] \times \left[\!\left[5,1,3\right]\!\right]$ with a MW 
$\left[\!\left[25,1,9\right]\!\right]$, 
we achieve a clear advantage with an approximately 6.25-fold improvement for a $10^{-2}$ depolarizing rate.




 We also compared these results with state of art codes using single logical qubit for comparable code rates and parameters (topological codes \cite{Kitaev_2003}), as depicted in Tab.~\ref{tab:table 1}. Our QBTC $\left[\!\left[5,1,3\right]\!\right] \times \left[\!\left[5,1,3\right]\!\right]$ code achieves consistently lower logical error rates than the surface code with the same code rate \cite{Forlivesi_2025} and generally outperforms the reinforcement learning method of toric codes \cite{Fitzek_2020} with distance $7$. However, the $XZZX$ surface code \cite{Bonilla_Ataides_2021} with distance $9$ demonstrates substantially better performance, achieving logical error rates approximately 50-fold lower than our code. This advantage is particularly pronounced under biased noise. As the noise becomes more balanced, the additional performance gain of the $XZZX$ code is reduced.

\begin{table}[b]
\centering
\vspace{-0.2cm}
\caption{Comparison of the performance of QECCs over 
a depolarizing channel : $\left[\!\left[25,1,9\right]\!\right]$ the concatenation of two $5$ qubit codes, QBTC $\left[\!\left[5,1,3\right]\!\right] \times \left[\!\left[5,1,3\right]\!\right]$ (worst case), $\left[\!\left[23,1,3/5\right]\!\right] \ A=10$ surface code simulation \cite{Forlivesi_2025}, Reinforcement learning method \cite{Fitzek_2020} and $XZZX$ surface code \cite{Bonilla_Ataides_2021}.}
\resizebox{\columnwidth}{!}{%
\begin{tabular}{c c c c c c}
\hline
 & $\left[\!\left[25,1,9\right]\!\right]$ & QBTC  & $\left[\!\left[23,1,3/5\right]\!\right]$  & RL $d=7$   & XZZX $d=9$  \\
\hline
$10^{-1}$ & $5\times10^{-2}$ & $3\times10^{-2}$ & $5\times10^{-2}$ & - & - \\
$10^{-2}$ & $10^{-5}$ & $1.6 \times 10^{-6}$ & $2\times10^{-4}$ & $4\times10^{-7}$ & - \\
$10^{-3}$ & $10^{-9}$ & $5.2 \times 10^{-11}$ & $10^{-6}$ & $8\times10^{-11}$ & $\simeq 10^{-12}$ \\
$10^{-4}$ & $10^{-13}$ & $5.3 \times 10^{-16}$ & - & $5\times10^{-15}$ & $\simeq 10^{-17}$ \\
\hline
\end{tabular}%
}

\vspace{2mm}

\label{tab:table 1}
\end{table}

After treating a specific case of QBTC that encodes only one logical qubit, we focus on more interesting combinations of constituent codes. Indeed, the QBTC $\left[\!\left[5,1,3\right]\!\right] \times \left[\!\left[5,1,3\right]\!\right]$  does not demonstrate the advantages of our turbo iterative decoding and we believe this to be the case because the exchange of extrinsic information might be limited with only one logical qubit per block. In order to clearly observe the turbo effect, we need thus to increase the number of logical qubits. 




\begin{figure}[t]
\vspace{-0.75cm}

\includegraphics[width=\linewidth]{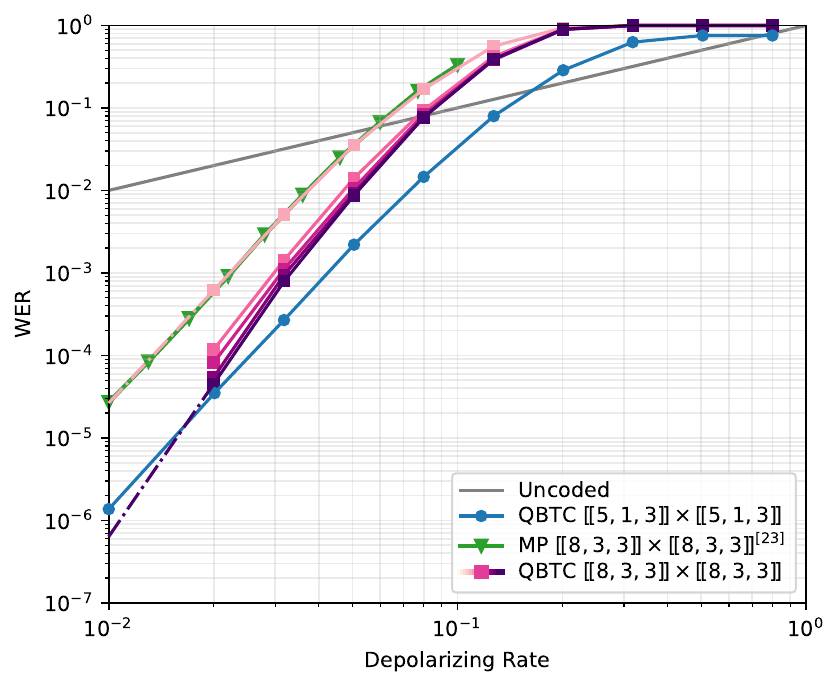}

\caption{Comparison performances of  QBTC $\left[\!\left[5,1,3\right]\!\right] \times \left[\!\left[5,1,3\right]\!\right]$, QBTC $\left[\!\left[8,3,3\right]\!\right] \times \left[\!\left[8,3,3\right]\!\right]$ (linear extrapolation at $10^{-2}$) and Poulin's message passing (MP) decoding \cite{Poulin_2006}.}
\vspace{-0.5cm}
\label{fig:833}
\end{figure}



Our next simulations use the 
QBTC $\left[\!\left[8,3,3\right]\!\right] \times \left[\!\left[8,3,3\right]\!\right]$ presented in 
Fig.~\ref{fig:833}, also denoted QBTC $\left[\!\left[64,9,9\right]\!\right]$, where darker colors correspond to higher decoder iteration indices. For depolarizing rates of $3.2\times10^{-2}$ and $2\times10^{-2}$, simulations were limited to 900, 253, 196, 167, 140 and 68, 13, 9, 6, 5 observed errors, respectively, over iterations $1$ to $5$. The QBTC 
$\left[\!\left[64,9,9\right]\!\right]$ performances clearly demonstrate graduate improvements at each iteration and illustrate the turbo effect.
We reach a WER of $6 \times 10^{-7}$ after 5 iterations by extrapolating for a depolarizing rate of $10^{-2}$. As a basis for comparison, we generalized the decoding method of  concatenated codes studied in \cite{Poulin_2006} used for codes with many logical qubits. This reference comparison was implemented with the assistance of an OpenAI GPT-5.5 model. We can observe that this reference decoder leads to the same results that we obtain for the first iteration of our QBTC, which demonstrates the advantage of our iterative decoder.

\begin{table}[b]
\centering
\vspace{-0.3cm}
\caption{Comparison of QBTC with QLDPC codes for a depolarizing rate of $10^{-2}$.}
\resizebox{\columnwidth}{!}{%
\begin{tabular}{c c c c c c}
\hline
code & code rate & phy qubits  &  $\rho_{l}$  & $\rho_{L}$   \\
\hline
QBTC $\left[\!\left[64,9,9\right]\!\right]$
 & $0,14$ & $7680$
 &  $6 \times 10^{-7}$ (5 iter)& $7.2 \times 10^{-5}$  \\
qLDPC $\left[\!\left[48,6,8\right]\!\right]$
 & $0,125$ & $8640$ &  $2 \times 10^{-4}$ (BP-OSD10) \cite{Panteleev_2021}  & $3.5\times10^{-2}$  \\
qLDPC $\left[\!\left[90,8,10\right]\!\right]$
 & $0,088$ &  $12150$ &  $3 \times 10^{-7}$ (AMBP4) \cite{kung2026efficientapproximatedegenerateordered} & $4\times10^{-5}$  \\
qLDPC $\left[\!\left[72,12,6\right]\!\right]$
 & $0,16$ & $6480$ &  $7 \times 10^{-5}$ (AMBP4) \cite{kung2026efficientapproximatedegenerateordered} & $6.3\times10^{-3}$  \\
 qLDPC $\left[\!\left[144,12,12\right]\!\right]$
 & $0,083$ & $12960$   & $3 \times 10^{-7}$ (BP-OSD0) \cite{iolius2024closedbranchdecoderquantumldpc} & $2.7\times10^{-5}$  \\
 qLDPC $\left[\!\left[126,28,8\right]\!\right]$
 & $0,22$ & $4536$ &  $3 \times 10^{-6}$ (BP) \cite{Panteleev_2021} & $1.17\times10^{-4}$  \\
 qLDPC $\left[\!\left[900,50,15\right]\!\right]$
 & $0,055$ & $19800$  &   $10^{-7}$ (BP-OSD10) \cite{Panteleev_2021}& $2.2\times10^{-6}$  \\
 
\hline
\end{tabular}%
}

\vspace{2mm}

\label{tab:table qLDPC}
\end{table}

In Tab.~\ref{tab:table qLDPC}, we collect different codes that encodes many qubits with different code rates and different word error rates $\rho_{l}$ for a depolarizing rate of $10^{-2}$. To illustrate and compare the scalability of different codes, we consider a use-case  where the electronic structure of a molecule is represented and simulated on a quantum computer
as an example \cite{Steudtner_2023}.
 We fix the number of needed logical qubits to $1080$ then we calculate the number of physical qubits required by using several blocks of the same code, alongside with the global word error rate:

\begin{equation}
\rho_{L}=1-(1-\rho_{l})^{\#blocks}
\end{equation}

Taking the example of our QBTC $\left[\!\left[64,9,9\right]\!\right]$ code, we would need $1080/9=120$ blocks leading to $120 \times 64 = 7680$ physical qubits and a global logical error rate of $1-(1-6 \times 10^{-7})^{120}=7.2 \times 10^{-5}$ with five iterations. Analyzing these results, we notice that qLDPC codes $\left[\!\left[90,8,10\right]\!\right]$ and $\left[\!\left[144,12,12\right]\!\right]$ that have $\rho_{L}$ values comparable to our code require nearly twice as much 
physical qubits whereas the $\left[\!\left[900,50,15\right]\!\right]$ code, which outperforms us, would require nearly three times more physical qubits for the same task, while also relying on the computationally demanding BP-OSD decoder. This illustrates the advantage of QBTCs: by using two small constituent blocks encoded in a product structure, one could create a code with a high code rate without requiring as much physical qubits as other QECCs. We can thus use this basic building block to encode more logical qubits while maintaining high performances.
In addition, our structure enables decoders to run in parallel for faster processing, while keeping each decoder’s qubits physically close together to simplify syndrome computation on the quantum processor.


 



Following these results, we expect to have more remarkable turbo effect by increasing the number of logical qubits with bigger constituent codes as discussed in Section.\ref{sec:morelog} or by using multi dimensional product codes as discussed in Section.\ref{sec:MLC}. These would imply, however, higher complexity per classical QBTC decoding block, which would need to be addressed at the hardware level.

It is worth noting that since we are dealing with stabilizer codes, we could use measurement circuits, as discussed in Section ~\ref{sec:BB}, to measure syndromes multiple times without going back to the logical level, paving the way to methods robust to noisy syndrome measurements. We believe that other fault-tolerant methods could also be used here such as cat states, flag error correction and Steane error correction\cite{Forlivesi_2025}.

\section{Perspectives}\label{sec:improv}

This section outlines potential improvements that could strengthen the current work and enhance its overall effectiveness. These would target more prominent turbo effect as well as practical directions for refinement that will be explored in future works.

\subsection{More Logical Qubits, Better Turbo Effect}\label{sec:morelog}

Based on the presented results, we believe that the key to a pronounced turbo effect lies in higher code parameters such as the number of  logical qubits and distance. This hypothesis will be further investigated and tested in future studies. 
A potential candidate would be the QBTC 
composed by the combination of two constituent $\left[\!\left[15,7,3\right]\!\right]$ $15$--qubit Hamming codes and with a global code rate of $0.21$. We could also increase the distance by using the product of two $\left[\!\left[15,3,5\right]\!\right]$ constituent codes to construct  a QBTC $\left[\!\left[225,9,25\right]\!\right]$ or by going up to the third dimension with three $\left[\!\left[15,7,3\right]\!\right]$ codes, enabling the creation of a QBTC $\left[\!\left[3375,343,27\right]\!\right]$ with distance $27$, a significant code rate of $0.1$ and a potentially strong turbo effect thanks to the high number of logical qubits. 
This demonstrates the flexibility of QBTCs with respect to variations in the code parameters, while extending the construction to higher dimensions and generalizing the theory of turbo iterative decoding, as explained in Section.~\ref{sec:MLC}.


\subsection{Mutli-layer Concatenation}\label{sec:MLC}

Recently, concatenated codes have received renewed interest, driven by advances in low-overhead fault-tolerant protocols using high-rate concatenated Hamming codes. These protocols can provide constant space overhead together with quasi-polylogarithmic time overhead \cite{Yamasaki_2024,Yoshida_2025}. Interest has also grown because of the development of high-rate many-hypercube (MHC) code families \cite{goto2024manyhypercube,xbzn-vn37,goto2026optimizedmanyhypercubecodeslower}. In high-rate concatenated codes, many logical qubits are obtained by arranging the inner and outer codes in parallel and interleaving their structure.

\begin{figure}[b]
\vspace{-0.4cm}
\centerline{\includegraphics[width=\linewidth]{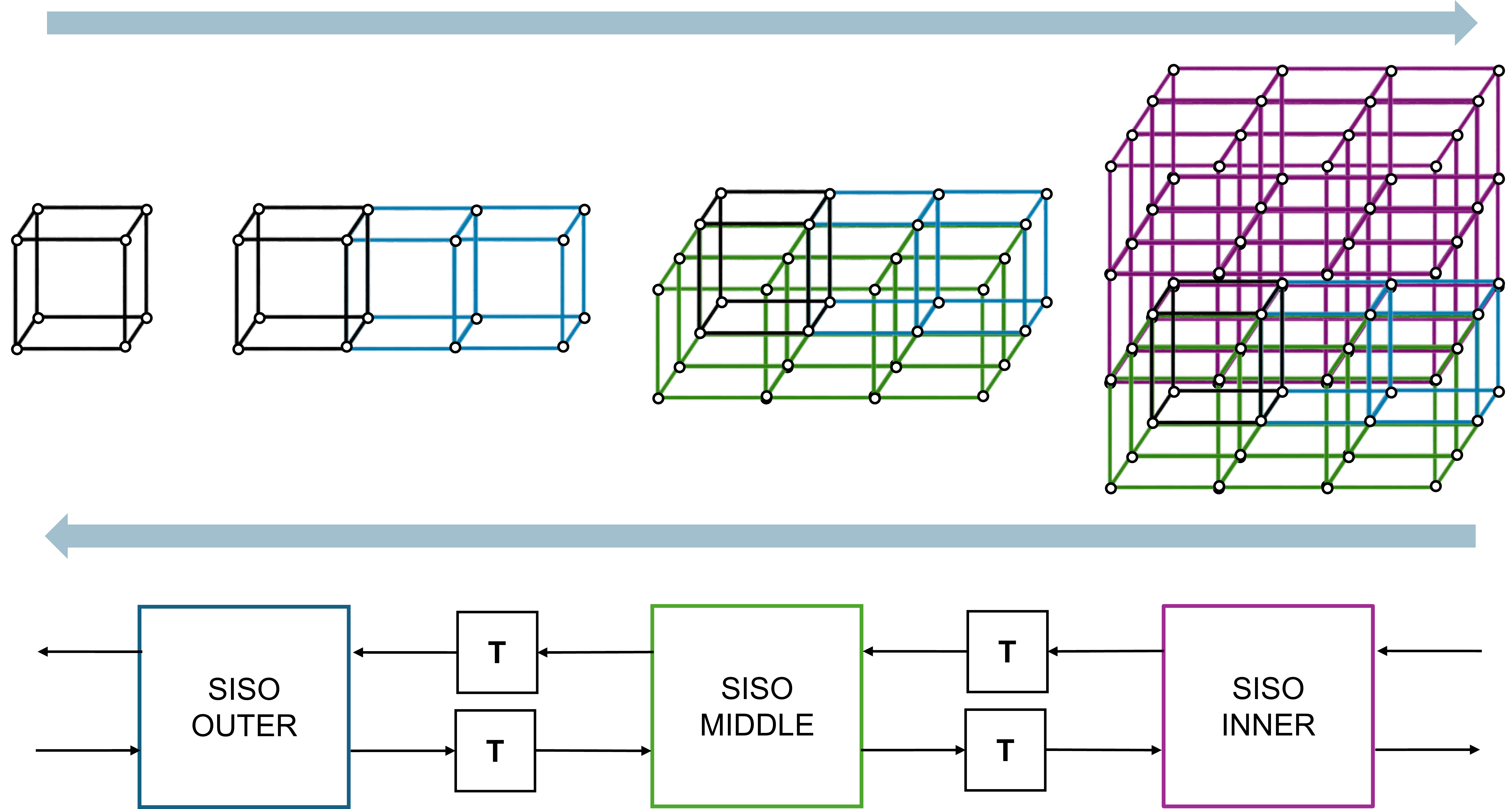}}
\caption{3-layer product code encoding and decoding of $\left[\!\left[4,2,2\right]\!\right]$ : Qubits are represented by vertices and physical and logical ports of siso units are respectively on the right and on the left.}
\label{fig:multi layer}
\end{figure}

We could extend the theory of QBTC to encode and decode 
multi-dimension product codes using multi layers structure as shown in Fig.~\ref{fig:multi layer}. The SISO units remain the same as for the 2D case presented here. The only difference is that we would need to add an extra SISO block for each layer between the two inner and outer decoders at the extremities. These blocks exchange soft information through transposition going up from inner to outer and then the other way around to ensure the iterative process. We believe that increasing the number of layers could make the turbo effect more significant while enabling higher distances.

\subsection{Physical-level Correction}\label{sec:BB}

The method described previously proposes a correction at the logical level. We utilize the inverse encoding scheme to access the syndromes. For a physical level correction using a QBTC, we could use measurement circuits since we use freely any stabilizer code.


\begin{figure}[t]
\vspace{-0.6cm}
\centerline{\includegraphics[width=\linewidth]{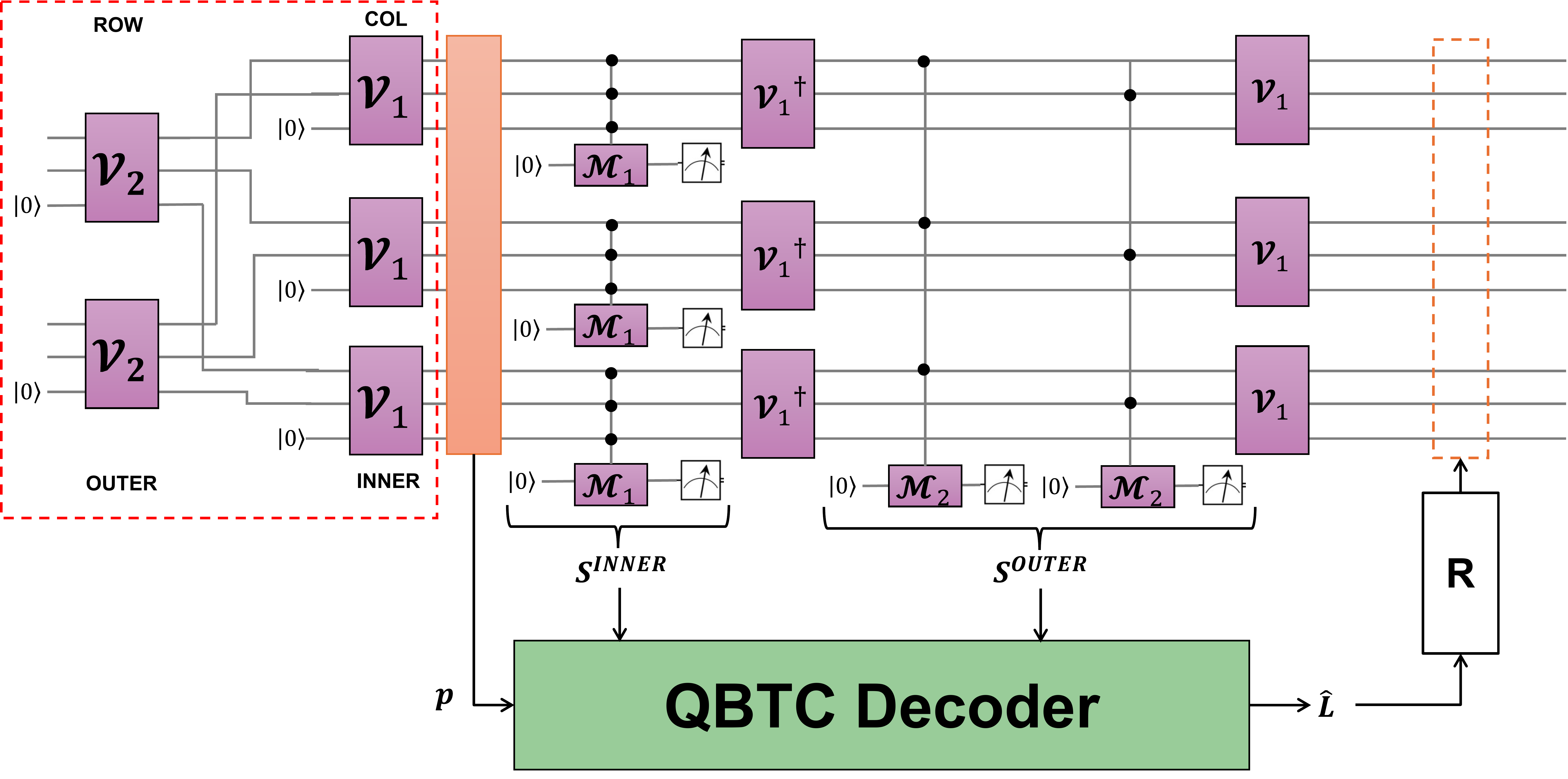}}
\caption{Example of syndrome extractions and error correction on the physical level for a QBTC $(3,2) \times (3,2)$.}
\label{fig:physical measurements}
\vspace{-0.5cm}
\end{figure}

By means of such measurement circuits, we can extract the syndromes 
without fully decoding back to the logical state. This allows us to preserve the complete encoded state while applying the necessary corrections directly at the physical level. The procedure is illustrated in Fig.~\ref{fig:physical measurements}.


In this case, we start with the QBTC encoding and we pass the encoded qubits through the noisy channel, just like before. However, we introduce extra ancilla qubits and we use the inner measurement circuit to measure the column syndromes. Then, we decode the inner code and use the outer measurement circuit to measure the row syndromes. Finally, we re-encode to come back to the full encoded state. Each of these row and column syndrome measurements can be done repeatedly by adding each time new ancilla qubits. This method is inspired by the syndrome extraction in the 9--qubit Shor code \cite{PhysRevA.52.R2493}. The syndromes along with the physical error rate are given to the classical QBTC decoder, which returns an estimate of the logical errors. In order to get a physical correction, we benefit from degeneracy. We take the logical estimate and ancilla errors candidates corresponding to the given syndromes and we encode them classically using the symplectic computation to obtain, at the end, an estimate of the physical errors. In addition, we could take a physical correction decision based on the physical output of the inner block.

\subsection{QBTC with Memory Qubits}

Instead of stabilizer codes, we could use the building blocks of quantum convolutional encoders. These use the transfer of memory qubits as presented in \cite{poulin2009quantumserialturbocodes}, with the restriction on the code parameters $k_1=n_2$ to guarantee the transfer of all memory qubits between the inner and outer blocks without cloning. 

\section{Conclusion}\label{sec:conc}


In this work, we propose and assess through simulations a new type of QECC, namely Quantum Block Turbo Codes, a type of product codes that use two or more constituent stabilizer codes. One of the greatest advantages of such solution is the flexibility provided on the combination of the constituent codes, allowing a fine trade-off between code-rate, number of logical qubits, code distance and decoding complexity.

We assessed here the simplest, 2D flavor of QBTCs, where this structure is used to exchange soft information while iteratively decoding rows and columns of qubits. We layed the foundations of our SISO decoder as well as the probabilistic transposition operations needed when passing information from one dimension to another during decoding.  
        
Our first evaluation of a QBTC $\left[\!\left[5,1,3\right]\!\right] \times \left[\!\left[5,1,3\right]\!\right]\equiv\left[\!\left[25,1,9\right]\!\right]$ with a single logical qubit has shown advantage compared to minimum weight decoding technique and demonstrates comparable results to decoding techniques for topological codes, but with no  turbo decoding effect. The code provided, however, good performances, with logical error rates of $1.6\times10^{-6}$ at $10^{-2}$ and estimated $5.2\times10^{-11}$ at $10^{-3}$ depolarizing rates. The use of constituent codes with higher number of logical qubits, namely with a QBTC $\left[\!\left[8,3,3\right]\!\right] \times \left[\!\left[8,3,3\right]\!\right]\equiv \left[\!\left[64,9,9\right]\!\right]$ provided a clear turbo decoding effect. We estimate for the fifth iteration a WER of $6\times10^{-7}$ at $10^{-2}$ and $10^{-12}$ at $10^{-3}$. 

The nature of QBTCs allows the construction of QECCs with a high code rate and very good performance due to turbo decoding effect and high achievable distance through the combination of constituent codes. Good performances are maintained even after using many basic QBTC blocks (for instance, several $\left[\!\left[64,9,9\right]\!\right]$) to encode a greater number of logical qubits while minimizing the required number of physical qubits compared to state of the art QECCs. Moreover, we believe that QBTC could potentially deliver stronger turbo effect by increasing both the number of logical qubits and the code distance. We can achieve this by using bigger constituent codes like $\left[\!\left[15,3,5\right]\!\right]$ and $\left[\!\left[15,7,3\right]\!\right]$ or by using multi dimensional ($>2$D) product codes. This will be addressed in future work. In addition, this method has a more practical use case by utilizing measurement circuits paving the way to fault tolerance in the future. Finally, the theory can be extended more generally to codes that use block constituent codes. 


\appendices

\section{Half iteration pseudo code}\label{halfiter}

\begin{algorithm}[t]
\small
\caption{Half iteration}
\label{alg:halfiter}
\begin{algorithmic}
\Require syndrome, logical prob., physical prob.
\Ensure logical prob. updated, physical prob. updated
\smallskip
\State \textbf{Logical update: } logical error $L$ 
\State synd candidates = quantumChase(syndrome)
\State $s=0$
\For{$S$ in synd candidates}
    \State $P=(L:S)U$
    \State $s \ +=$ physical prob.($P$)
\EndFor
\State logical prob. updated($L$) = logical prob.($L$) * $s$
\smallskip
\State \textbf{Physical update: } physical error $P$
\State synd candidates = quantumChase(syndrome)
\State logical errors = Pauli group($k$)
\State $s=0$
\For{$S$ in synd candidates}
    \For {$L$ in logical errors}
        \If{$P==(L:S)U$}
            \State $s \ +=$ logical prob.($L$)
        \EndIf
    \EndFor
\EndFor
\State physical prob. updated($P$) = physical prob.($P$) * $s$
\end{algorithmic}
\end{algorithm}

\begin{algorithm}[t]
\small
\caption{Transposition}
\label{alg:training}
\begin{algorithmic}
\smallskip
\State \textbf{Marginalize: }
\For{each position $(i,j)$}
    \For{$C$ in columns}
        \If{$C[i]=I$}
            \State $\mathbb{P}_{i,j}(I)+=\mathbb{P}_j(C)$
        \EndIf
        \If{$C[i]=X$}
            \State $\mathbb{P}_{i,j}(X)+=\mathbb{P}_j(C)$
        \EndIf
        \If{$C[i]=Y$}
            \State $\mathbb{P}_{i,j}(Y)+=\mathbb{P}_j(C)$
        \EndIf
        \If{$C[i]=Z$}
            \State $\mathbb{P}_{i,j}(Z)+=\mathbb{P}_j(C)$
        \EndIf
    \EndFor
\EndFor
\smallskip
\State \textbf{Reorganize: }
\For{$L$ in rows}
    \State $\mathbb{P}_i(L)=\prod_{j}\mathbb{P}_{i,j}(L[j])$
\EndFor
\end{algorithmic}
\end{algorithm}

\begin{figure}[t]
\vspace{-0.6cm}

\centerline{\includegraphics[width=0.7\linewidth]{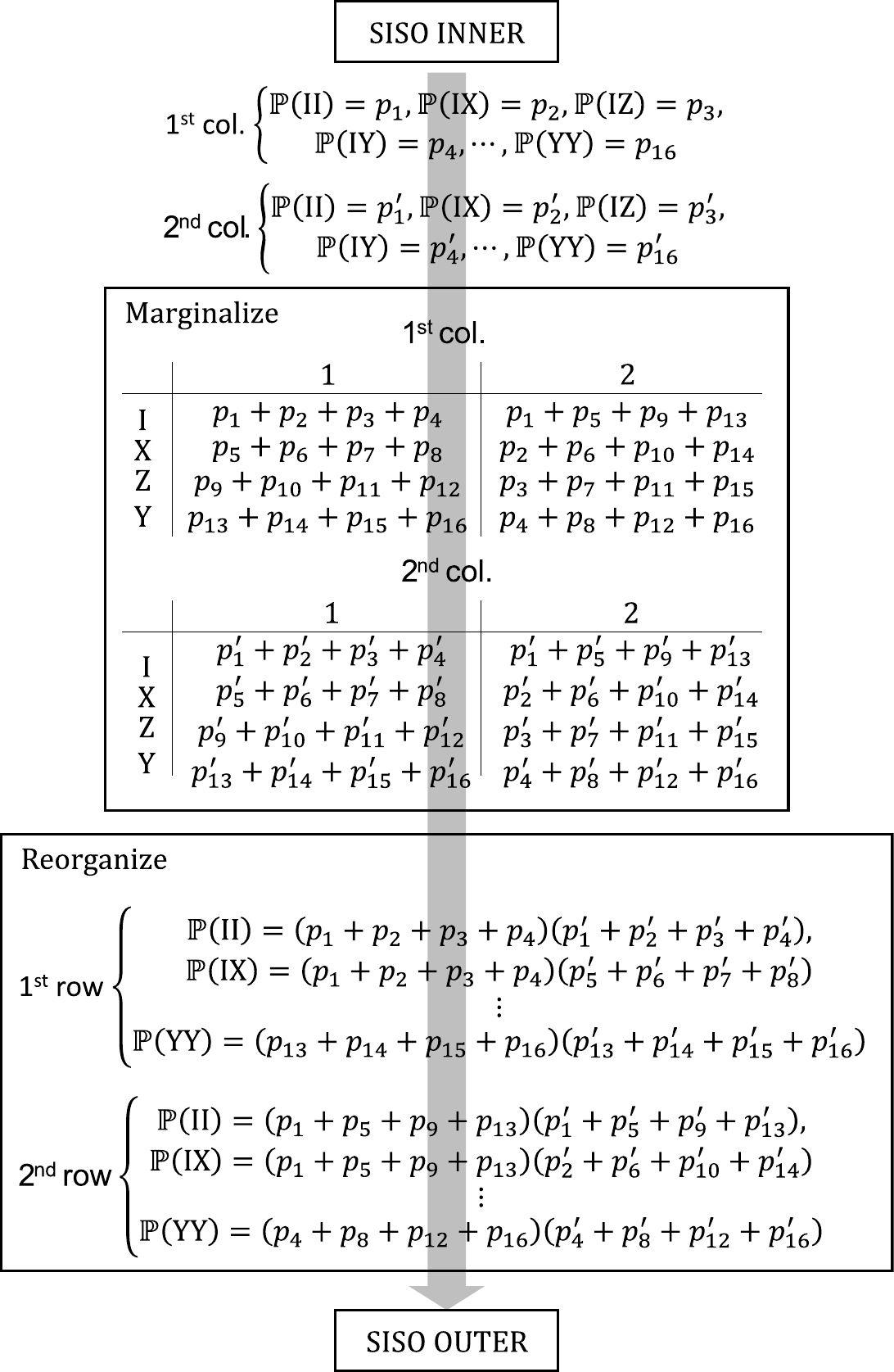}}
\caption{Inner-outer transposition example for QBTC $(3,2) \times (2,1).$}
\label{fig: transpose}
\vspace{-0.6cm}
\end{figure}

\vspace{-0.1cm}
Algorithm \ref{alg:halfiter} presents a pseudo code for the half iteration described in Section.\ref{sec:decoding}. Many features in the following algorithm could be optimized and parallelized to reduce its complexity. For example, when receiving the syndromes, we could calculate once and store not only all the physical candidates for all logical errors, needed in the logical update but also the logical candidates that could lead to physical errors, needed in the physical update. In addition, this algorithm could run for all columns or rows in parallel at each half-iteration. We could also reduce the complexity by limiting the weights of the inner physical candidates while maintaining the same level of performance.


\section{Transposition pseudo code}\label{Trans}


Algorithm \ref{alg:training} shows the pseudocode for the M \& R procedure described in Section.\ref{sec:MR}. We could easily determine the Pauli sequences that have $I,X,Y,Z$ in a certain position following the pattern of generation of the sequences without needing to go through all of them. This pseudo code is just an illustration of the general idea of probabilistic transposition.




\bibliographystyle{IEEEtran}
\bibliography{references}

\end{document}